# Compton suppression in BEGe detectors by digital pulse shape analysis

Yu-Hao Mi[a]   Hao Ma[a]   Zhi Zeng[a*]   Jian-Ping Cheng[a]   Jun-Li Li[a]   Hui Zhang[a]

[a] Key Laboratory of Particle and Radiation Imaging (Ministry of Education) and Department of Engineering Physics, Tsinghua University, Beijing 100084, China

Abstract: A new method of pulse shape discrimination (PSD) for BEGe detectors is developed to suppress Compton-continuum by digital pulse shape analysis (PSA), which helps reduce the Compton background level in gamma ray spectrometry. A decision parameter related to the rise time of a pulse shape was presented. The method was verified by experiments using $^{60}$Co and $^{137}$Cs sources. The result indicated that the $^{60}$Co Peak to Compton ratio and the Cs-Peak to Co-Compton ratio could be improved by more than two and three times, respectively.
Key words: PSA, PSD, Compton suppression, BEGe

1. Introduction

Background suppression techniques are important for applications with germanium detectors, such as gamma ray spectrometry and rare event detection. Passive and active shields are traditionally used to achieve a lower background level for these applications (Heusser, 1995; Hult, 2007). In the last 30 years, pulse shape analysis (PSA) has attracted more attention with the fast development of electronics and computers. Lots of efforts have been spent on developing PSA techniques for background suppression of germanium detectors and PSA becomes an alternative or complement to traditional techniques (Aalseth et al., 2013; Agostini et al., 2013; Akkoyun et al., 2012; Aspacher et al., 1992; Barbeau et al., 2007; Cooper et al., 2011; Crespi et al., 2009; Crespi et al., 2008; Crespi et al., 2007; Elliott et al., 2006; Feffer et al., 1989; Gonzalez et al., 2003; Hellmig and Klapdor-Kleingrothaus, 2000; Petry et al., 1993; Philhour et al., 1998; Roth et al., 1984; Schmid et al., 1999; Yue et al., 2014). As to PSA of germanium detectors, pulse shapes from the pre-amplifier or timing filter amplifier are usually recorded by the FADC with a high sampling rate and analyzed offline to extract useful information. For background suppression, differences between pulse shapes of background and the concerned signal must be determined to further discriminate background events, of which the process is called pulse shape discrimination (PSD).

Many PSD techniques of semi-coaxial germanium detectors have been developed for different purposes, such as Compton background suppression in conventional gamma ray spectrometry (Aspacher et al., 1992; Schmid et al., 1999), beta background suppression in gamma ray astrophysics (Feffer et al., 1989; Petry et al., 1993; Philhour et al., 1998) and multi-site event (MSE) background suppression in neutrinoless double beta decay (0νββ) detection (González et al., 2003; Hellmig and Klapdor-Kleingrothaus, 2000). Some of these techniques are based on decision parameters extracted through direct analysis of raw pulse shapes, while others are based on comparison with standard pulse shapes generated by experiments or simulations. Aspacher et al. adopted the fall time of current pulses as a decision parameter to suppress the Compton continuum by digital PSA and improved the Cs-Peak to Co-Compton (Cs-P/Co-C) ratio, i.e. the ratio of the maximum channel count in the 0.661 MeV gamma peak region to the average channel count of the Compton-continuum of $^{60}$Co's gamma rays, by a factor of two (Aspacher et al., 1992). However,

---

the PSD power of semi-coaxial germanium detectors is restricted by the relatively uniform distribution of the electric filed inside the germanium crystal.

Broad energy germanium (BEGe) detectors are comparatively more powerful in PSD due to the much smaller inner electrode and the consequent sharply varying inner electric field. Thereby scientists used BEGe detectors in 0νββ experiments to efficiently discriminate MSE background. For instance, GERDA collaboration proposed the "A/E" technique for the discrimination of single-site events (SSE) and MSEs (Budjáš et al., 2009). González De Orduña et al. used the "A/E" technique to suppress the Compton continuum as well and improved the $^{60}$Co Peak to Compton (P/C) ratio by a factor of 1.6 ~ 2 (González De Orduña et al., 2010).

In this work, we present a new PSD method for Compton suppression in BEGe detectors. The rise time of charge pulses from the pre-amplifier was concerned and the corresponding decision parameter was presented. Experiments with $^{60}$Co and $^{137}$Cs sources were carried out to study this method. Preliminary simulation study, including pulse shape simulations (PSS) and Monte Carlo (MC) simulations, was conducted to explore the mechanism of this method.

2. Experimental setup and data acquisition

A Canberra p-type BEGe detector (BE6530) integrated with a charge-sensitive pre-amplifier (2002C) was used in this study. The germanium crystal is 91.1 mm in diameter and 31.4 mm in height, and the inner electrode is 13.5 mm in diameter. During the experiments, the detector was located aboveground and operated without passive shields. A 250MS/s 12 bit CAEN digitizer (DT5720E) was used to record pulses (32 us long) when two point-like sources of $^{60}$Co and $^{137}$Cs respectively were located ~7.5 cm above the top surface meanwhile. The sources are sealed in the thin plastic film (Φ 32 mm × 4 mm) and their radioactivity are in the level of several thousand Bq. Totally $2\times10^6$ pulse shapes were acquired.

Digitized pulse shapes from the pre-amplifier were treated as raw data and processed offline by PSA techniques directly while no analog electronics were further used. First, a simple selection was applied to the pulse shapes by checking the average value and linear slope of the baseline. Second, a pile-up rejection algorithm was applied so that the superposition of several pulse shapes was identified and rejected. The above two steps were intended for the basic quality assurance of the raw data. Third, all selected pulse shapes were baseline-restored and either processed by the trapezoidal shaping algorithm (Jordanov and Knoll, 1994) to extract the energy information of the corresponding physical event or analyzed for the purpose of PSD.

3. PSD method

The rise part of pulse shapes from the pre-amplifier corresponds to the drift of electrons and holes inside the crystal and accordingly contains information about the interaction of the physical event, such as the position of the interaction and the total interaction number. Thus, it is feasible to extract such information by analyzing the rise part in a PSD process.

In this study, a smaller rise time parameter and $r_{90}$ of the pulse shape are concerned and we take the example of $r_{50}$ and $r_{90}$. $r_{50}$ is the rise time between 10% and 50% of the maximum amplitude of the pulse shape, while $r_{90}$ is the one between 10% and 90%. Figure 1 is a two-dimension scatter plot showing the joint distribution of pulse shapes' $r_{50}$ and $r_{90}$. The dots seem to concentrate around two "lines": the inclined line A and the horizontal line B. Line A represents pulse shapes of which the ratio of $r_{50}$ to $r_{90}$ is large and tend to be constant, while Line B represents those of which the ratio of $r_{50}$ to $r_{90}$ is small and varying. Figure 2 (a) shows a typical pulse shape corresponding to

Line A, which rises slowly at first but then sharply to the maximum amplitude. Such pulse shapes mainly correspond to events interacting far from the inner electrode, and the initial collection of charge mainly occurs in the region with weak electric field where charge carriers travel slowly. Figure 2 (b) shows another corresponding to Line B, which rises fast at first but then slowly. Such pulse shapes mainly correspond to events interacting close to the inner electrode, and the initial collection of charge mainly occurs in the region with strong electric field where charge carriers travel fast.

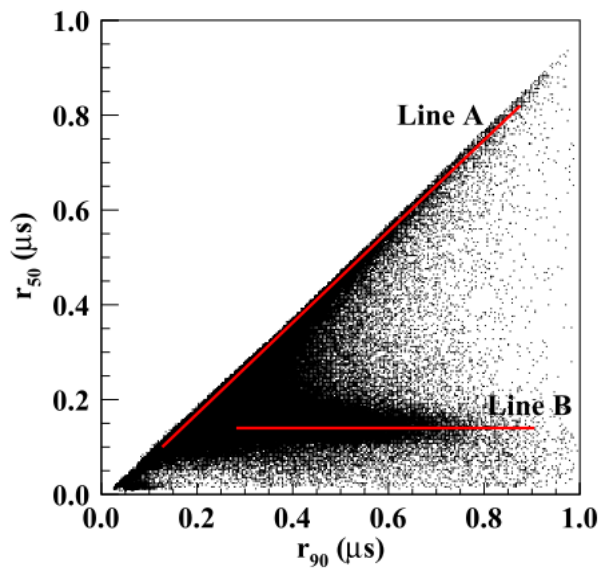

Figure 1 The scatter plot of $r_{50}$ vs $r_{90}$.

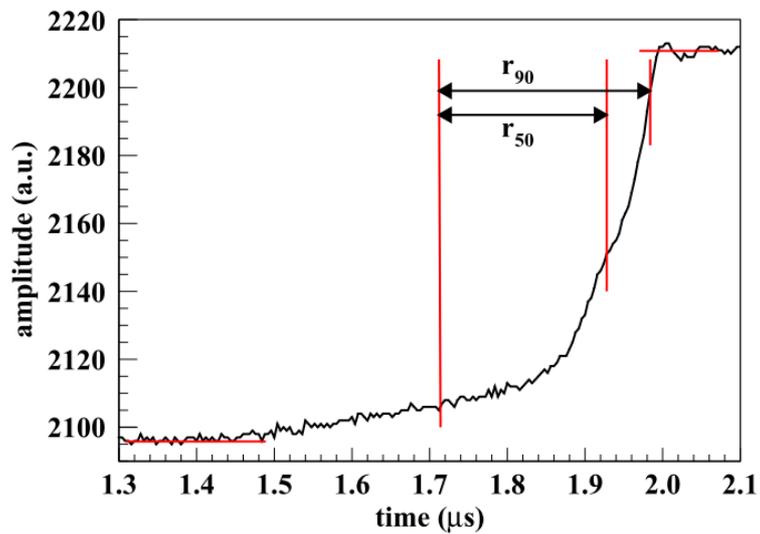

(a)

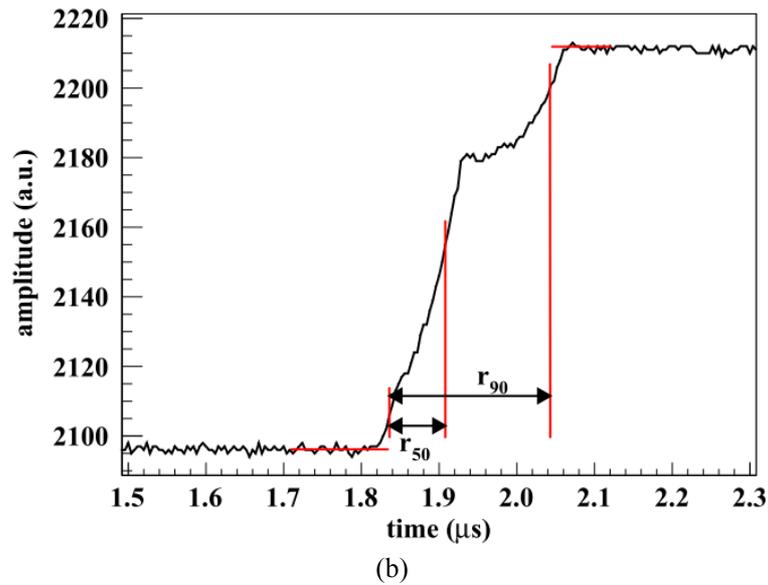

(b)

Figure 2 The typical pulse shapes corresponding to: (a) Line A of which the ratio of $r_{50}$ to $r_{90}$ is large; (b) Line B of which the ratio of $r_{50}$ to $r_{90}$ is small.

Based on the above analysis, a new parameter, $R_{59}$, defined as the ratio of $r_{50}$ to $r_{90}$ is presented. To some extent, $R_{59}$ indicates how far the event is interacting from the inner electrode. Figure 3 is the density plot of the $R_{59}$ distribution as a function of the event energy. Events in the Compton-continuum tend to concentrate around a horizontal band corresponding to a large $R_{59}$ value, while the peak events of gamma rays from $^{60}$Co tend to be distributed in a vertical band corresponding to varying $R_{59}$ values in a wide range. The $R_{59}$ distributions for Compton events and peak events are notably different, thus $R_{59}$ can be used as the decision parameter to discriminate Compton events from peak events, i.e. to suppress the Compton-continuum. As shown in Figure 3, the cut threshold can be set as the lower boundary of the horizontal band. Events with $R_{59}$ values larger than the threshold will be rejected, while the others will be accepted. It is worth mentioning that this PSD method will cut some fraction of the peak events and keep some Compton events as well, which is not the ideal condition. However, the peak is still more prominent after the cut, which is confirmed in Chapter 4.1.

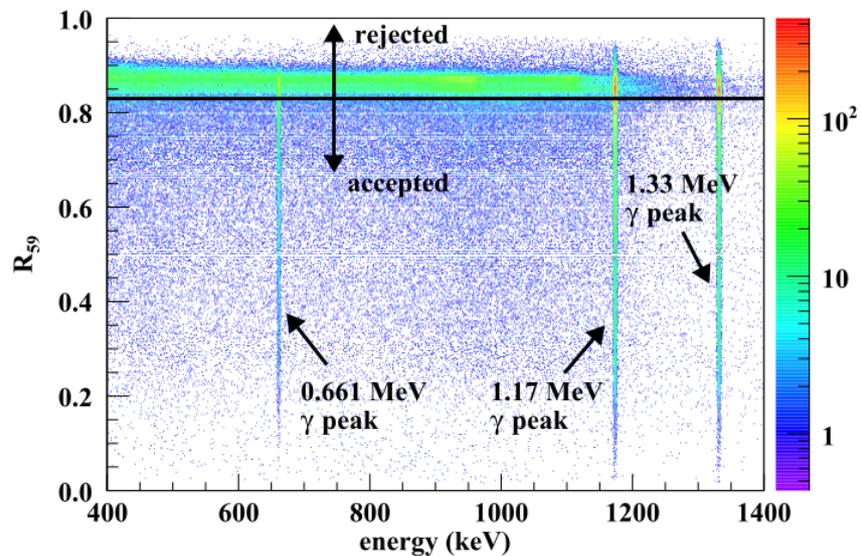

Figure 3 The density plot of the $R_{59}$ distribution as a function of the event energy. The color bar indicates in logarithmic scale the number of events in a square of 1 keV × 0.0025. The black line shows a possible cut threshold.

Actually, the decision parameter is calculated as the ratio of a smaller rise time parameter to $r_{90}$, and the smaller one can be changed to $r_x$ which corresponds to the rise time between 10% and x% of the maximum amplitude of the pulse shape. Analysis similar to those above in this chapter can be deduced when other $r_x$ values are used, but the Compton suppression effect will deteriorate when $r_x$ is too large. The results of Compton suppression using different $r_x$ values are shown in Chapter 4.2.

4. Results and discussions

4.1. Compton suppression results using $r_{50}$

According to the distribution in Figure 3, the cut threshold was set as 0.83 and the energy spectra before and after the cut are shown in Figure 4. The P/C ratio of $^{60}$Co, i.e. the ratio of the maximum channel count in the 1.33 MeV gamma peak region to the average channel count of the Compton-continuum between 1.040 MeV and 1.096 MeV, was calculated for both spectra in Figure 4 and turned out to be 34.0 and 81.9 before and after the cut respectively. Figure 5 shows the dependency of the $^{60}$Co P/C ratio, the 1.33 MeV gamma peak area and the efficiency on the cut threshold of $R_{59}$ and all data have been normalized to the corresponding values of the original spectrum. The efficiency, defined as the product of the $^{60}$Co P/C ratio and the square root of the 1.33 MeV gamma peak area, is related to the statistical property of the spectrum (Aspacher et al., 1992). The P/C ratio can be improved by more than 2 times when the cut threshold is set between 0.3 and 0.84, but the peak area is reduced more severely when the threshold becomes smaller. Thus, the cut threshold should be set as large as possible to maintain a good statistical property of the energy spectrum on the premise that the Compton suppression effect satisfies the demands.

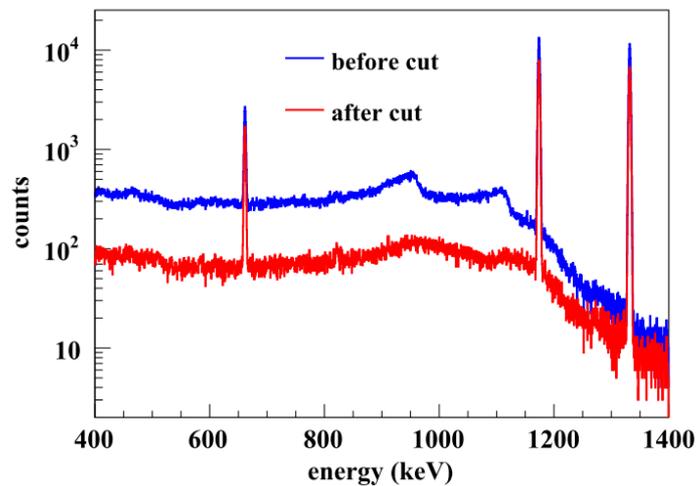

Figure 4 The energy spectra before and after the $R_{59}$ cut with a 0.83 threshold.

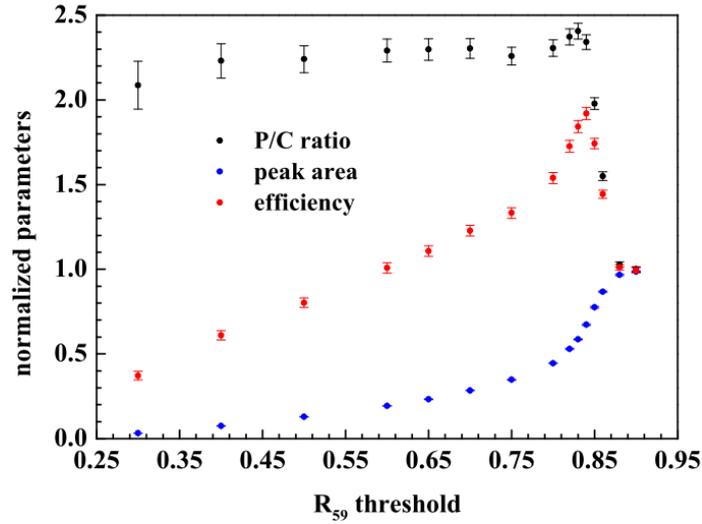

Figure 5 The dependency of the normalized $^{60}$Co P/C ratio, 1.33 MeV gamma peak area and efficiency on the $R_{59}$ threshold.

The Cs-P/Co-C ratio, for which the Compton-continuum is restricted between 0.680 MeV and 0.730 MeV, was also calculated for both spectra in Figure 4 and turned out to be 8.4 and 27.0 before and after the cut respectively. Figure 6 shows the dependency similar to that in Figure 5, where the Cs-P/Co-C ratio, the 0.661 MeV gamma peak area and the corresponding efficiency were concerned instead. The Cs-P/Co-C ratio can be improved by more than 3 times while the efficiency maintains large.

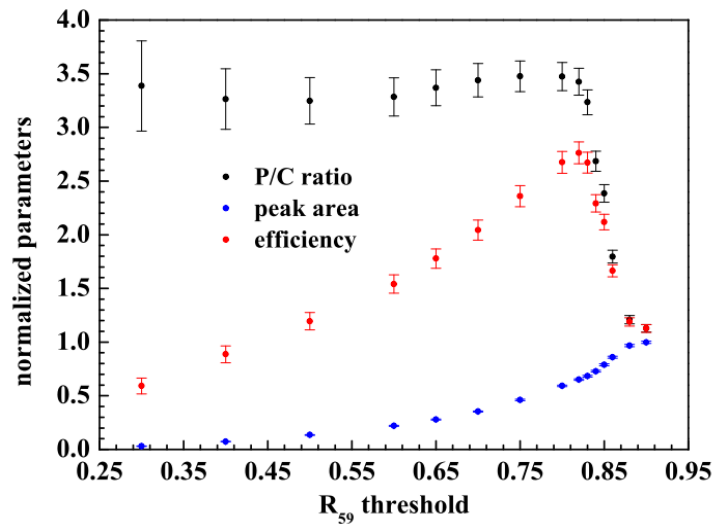

Figure 6 The dependency of the normalized Cs-P/Co-C ratio, 0.661 MeV gamma peak area and efficiency on the $R_{59}$ threshold.

4.2.  Compton suppression results using different $r_x$ values

The PSD processes based on different $r_x$ values respectively were applied to the pulse shapes and a typical cut threshold was used in each PSD process to obtain its representative result, as listed in Table 1. The $^{60}$Co P/C ratio is improved by more than 2 times in all conditions, but the efficiency begins to drop quickly when $r_x$ is larger than $r_{50}$. Thus, the PSD method is steady in performance when $r_x$ varies from $r_{20}$ to $r_{50}$, but the statistical property of the spectrum after the cut will deteriorate quickly when $r_x$ is larger than $r_{50}$.

Table 1 The Compton suppression results when PSD processes with different $r_x$ values were applied. The parameters in the last 3 columns were normalized to the corresponding values of the original spectrum.

| $r_x$ | decision parameter | cut threshold | norm. $^{60}$Co P/C ratio | norm. 1.33 MeV peak area | norm. efficiency |
|---|---|---|---|---|---|
| $r_{20}$ | $R_{29}$ | 0.49 | 2.31±0.04 | 0.6256±0.0034 | 1.827±0.035 |
| $r_{30}$ | $R_{39}$ | 0.67 | 2.43±0.05 | 0.6579±0.0036 | 1.967±0.037 |
| $r_{40}$ | $R_{49}$ | 0.77 | 2.39±0.04 | 0.6542±0.0036 | 1.936±0.036 |
| $r_{50}$ | $R_{59}$ | 0.83 | 2.41±0.05 | 0.5860±0.0033 | 1.842±0.016 |
| $r_{60}$ | $R_{69}$ | 0.88 | 2.21±0.04 | 0.5500±0.0032 | 1.640±0.032 |
| $r_{70}$ | $R_{79}$ | 0.92 | 2.05±0.04 | 0.4845±0.0029 | 1.424±0.029 |
| $r_{80}$ | $R_{89}$ | 0.95 | 2.11±0.05 | 0.4023±0.0026 | 1.339±0.029 |

4.3. Simulation studies

To study the mechanism of this PSD method, PSS and MC simulations were carried out. Figure 7 shows the geographical distribution of the decision parameter $R_{59}$ deduced from simulated pulse shapes corresponding to events interacting at different positions inside the germanium crystal, and the two-dimension map corresponds to half a cross section along the crystal's height and radius directions. The closer to the inner electrode the event interacts, the smaller the $R_{59}$ is. Besides, the events interacting in the large region far from the inner electrode tend to have a large and almost constant $R_{59}$ value. This varying tendency of $R_{59}$ is coincident with the description in Chapter 3. However, the absolute $R_{59}$ values obtained from PSS may not be coincident with the real values obtained from experiments, resulting from some estimated parameters of the detector instead of exclusive ones used in PSS.

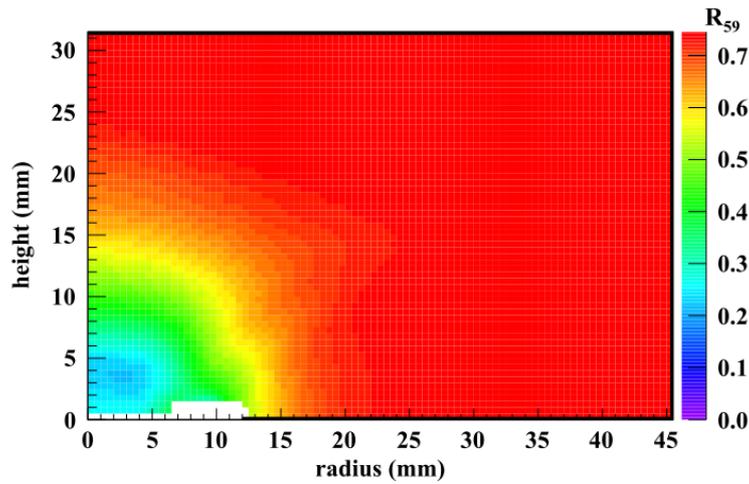

Figure 7 The geographical distribution of the decision parameter $R_{59}$ in the germanium crystal. The $R_{59}$ values were obtained from PSS with a 0.5 mm step length inside the crystal and indicated by the color bar.

In the MC simulation, the condition of the experiment was simulated. The probability distributions of the interacting positions of the 1.33 MeV peak events and the Compton events between 500 keV and 1000 keV are shown in Figure 8 (a) and Figure 8 (b) respectively, and the two-dimension map also corresponds to half a cross section as that in Figure 7. Compton events tend to interact in the peripheral region near the border of the crystal where $R_{59}$ values are larger, while peak events tend to interact deeper into the inner region of the crystal where $R_{59}$ values are smaller. This tendency implies the basic mechanism of this PSD method and it is the same with the processes using different $r_x$ values.

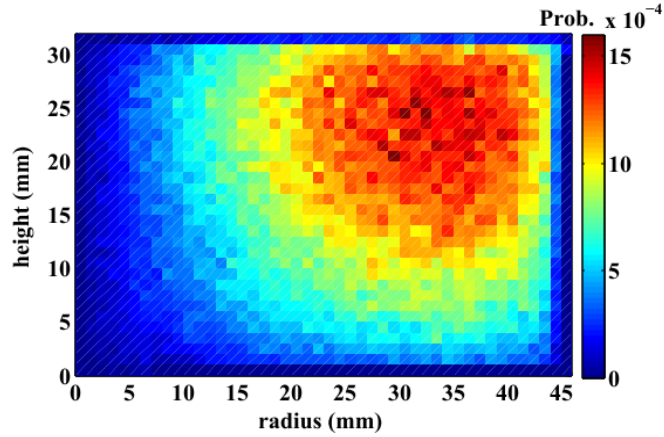

(a)

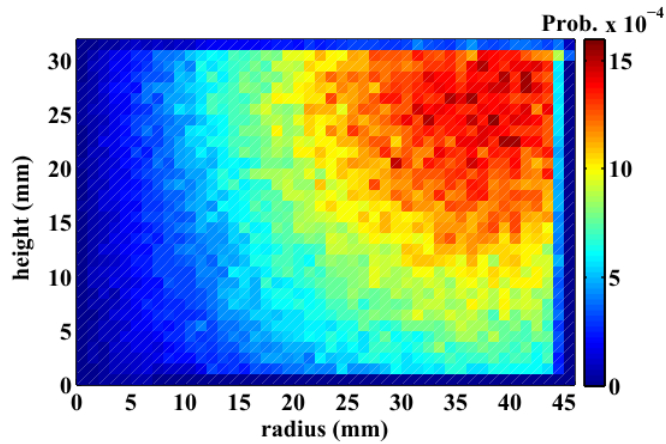

(b)

Figure 8 The probability distributions of the interacting positions of: (a) the 1.33 MeV peak events; (b) the Compton events between 500 keV and 1000 keV. The color bar indicates the probability that the corresponding events interact in a square of 1 mm × 1 mm inside the germanium crystal.

5. Conclusions

A new PSD method was developed to suppress Compton-continuum for BEGe detectors. Raw pulse shapes from the preamplifier were digitized, recorded by digitizers with a high sampling rate and analyzed offline by PSA techniques directly. By using the ratio of a smaller rise time $r_x$ to $r_{90}$ as the decision parameter in the PSD process, the $^{60}$Co P/C ratio and the Cs-P/Co-C ratio could be improved by more than two and three times, respectively. The PSD method's performance keeps good when $r_x$ varies from $r_{20}$ to $r_{50}$, but turns worse quickly when $r_x$ becomes larger due to the severe loss of peak area. PSS and MC simulations verify that Compton and peak events tend to interact near the border of the crystal and deeper into the crystal respectively, and the values of the decision parameter in the two regions are different, which implies the basic mechanism of the PSD method.

This PSD method is effective in the Compton suppression of gamma ray spectrometry and helps reduce the background level in low-background spectrometry.

Acknowledgements

This work is supported by National Natural Science Foundation of China (No.11175099 & No.11675088) and Tsinghua University Initiative Scientific Research Program (No.20151080354 & No.2014Z21016). We are grateful to the colleagues of CDEX collaboration for their help in the experiments.

contact germanium detector at the China Jingping Underground Laboratory. Phys Rev D, 90(9): 091701.